# Zero Point Entropy in Stuffed Spin Ice

G.C. Lau[1], R.S. Freitas[2], B.G. Ueland[2], B.D. Muegge[1], E.L. Duncan[1], P. Schiffer[2], and R.J. Cava[1]

[1]*Department of Chemistry, Princeton University, Princeton NJ 08544*

[2]*Department of Physics and Materials Research Institute, Pennsylvania State University, University Park, PA 16802*

The third law of thermodynamics dictates that the entropy of a system in thermal equilibrium goes to zero as its temperature approaches absolute zero. In ice, however, a "zero point" or residual entropy can be measured – attributable to a high degeneracy in the energetically preferred positions of the hydrogen ions associated with the so-called "ice rules".[1,2] Remarkably, the spins in certain magnetic materials with the pyrochlore structure of corner-sharing tetrahedra, called "spin ice", have an equivalent degeneracy of energetically preferred states and also have been shown to display a zero point entropy.[3,4,5,6,7] Here we report that we have chemically altered $Ho_2Ti_2O_7$ spin ice by stuffing extra Ho magnetic moments into normally non-magnetic Ti sites surrounding the Ho tetrahedra. The resulting series, $Ho_2(Ti_{2-x}Ho_x)O_{7-x/2}$, provides a unique opportunity to study the effects of increased connectivity between spins on a frustrated lattice. Surprisingly, the measured zero point entropy per spin appears unchanged by these excess spins, and the dynamic freezing of the spins is suppressed to lower temperatures. The results challenge our understanding of the spin ice state, and suggest a new avenue for using chemistry to study both ice-like frustration and the properties of the broad family of geometrically frustrated magnets based on the pyrochlore structure.

The energy equivalence of different configurations for hydrogen ions in ice is an example of geometrical frustration. In general, geometrical frustration occurs when local interactions between a system's components cannot all be simultaneously satisfied due to their spatial arrangement on a regular lattice, leading to a large degeneracy of low energy states. Geometrical frustration in magnetic materials[8,9] has led to a wide range of novel physics and has received particular attention in the rare earth pyrochlore compounds. These materials have the generic formula $R_2M_2O_7$, where the $R^{3+}$ and $M^{4+}$ cations form equivalent interpenetrating, equal size sublattices of corner-sharing tetrahedra (as shown in Fig. 1).[10] These systems display low temperature behavior ranging from exotic ordered states,[11,12,13,14] to spin liquids, in which the spins are correlated yet continue to fluctuate as $T \rightarrow 0$,[15] to the spin ice state.

Spin ice behavior has been observed in pyrochlores where the R site is occupied by either Dy or Ho magnetic cations, and non-magnetic Ti or Sn occupy the equivalent M sublattice.[5,6] The crystal field environment in this structure is such that each Ho or Dy spin must point along the line connecting the centers of its two tetrahedra (i.e., either directly toward or away from the centers of the tetrahedra). Coupled with the dipolar and exchange interactions, this anisotropy results in six energetically equivalent low-energy states for each tetrahedron, in which two spins point in to and two spins point out from each tetrahedron, exactly analogous to the 'ice rules' for hydrogen positions in water ice. Neutron scattering and magnetic susceptibility measurements have demonstrated that these systems freeze into a disordered spin state at low temperatures, and that the local spin correlations follow the two-in/two-out structure.[16,17,18,19] The statistical mechanics of this system is identical to that of the hydrogen ions in ice, and the measured molar entropy of the spin system (obtained by integrating the measured specific heat) is $S = R[\ln 2 - \frac{1}{2}\ln(^3/_2)]$ where $R$ is the gas constant.[4] This is less than the $S = R\ln 2$ value expected for an Ising system by precisely the same amount as for ice and can be understood as a direct consequence of the "ice rules" as applied by Pauling.[2]

The tetrahedra on the R and M sublattices of the pyrochlore spin ice materials have sides of the same length, i.e., the distance between nearest neighbors within each of the two sublattices is the same. Furthermore, the shortest distance between neighboring R and M sites is also exactly the same as this nearest neighbor distance within each of the two sublattices. As illustrated in Fig. 1a, this implies that if one were to consider all of the R and M sites together as a single lattice, the combined lattice would consist of side sharing tetrahedra (equivalent to a face-centered-cubic lattice, which also can lead to geometrical frustration[20,21]). The geometry of the two sublattices also implies that if one were to replace a single non-magnetic M ion with a magnetic R ion, the connectivity of the surrounding R sites would be drastically changed. Instead of having only six nearest neighbor spins, such a replacement would cause the adjacent R-site spins to have a seventh nearest neighbor spin. Furthermore, the spins on those adjacent R-site tetrahedra would share more than three nearest neighbors, which would be expected to have a substantial impact on the nature of the frustration.

We have explored the consequences of such an increase in connectivity by replacing a fraction of the non-magnetic Ti ions with Ho ions in $Ho_2Ti_2O_7$, effectively stuffing extra spins into the spin ice system. The resultant materials form the series $Ho_2(Ti_{2-x}Ho_x)O_{7-x/2}$, where $0 \leq x \leq 0.67$ (synthesis details are given in the Methods section). With the extra Ho, the magnetic sites are no longer limited to the vertices of corner-sharing tetrahedra and can have as many as 12 nearest neighbors (with the average number being $n = 6 + 3x$). Structure refinements revealed that for $0 \leq x \leq 0.3$, Ho replaced Ti strictly on the pyrochlore M-site. Above $x \cong 0.4$, the mixing of Ho and Ti is no longer restricted to the M site, and some Ti is situated on the R site, creating a fraction of disorder in the previously pure Ho corner sharing sublattice (Fig. 1b). This mixing of ions on the two sites increases with $x$ until Ho and Ti are completely randomized, and the R and M sites become indistinguishable from each other at $x = 0.67$ (i.e., $Ho_2TiO_5$). The $x = 0.67$ end member has the fluorite crystal structure – simply a pyrochlore with both R and M sites occupied randomly by Ho and Ti ions in their stoichiometric ratio.[22,23,24] The Ho doping of $Ho_2Ti_2O_7$ thus transforms separate networks of Ho and Ti corner-sharing tetrahedra into a network of edge-sharing tetrahedra where Ho and Ti are randomly mixed in a 2:1 ratio

The stuffing of $Ho_2Ti_2O_7$ offers a unique opportunity to study the nature of ice-like frustration, since the additional Ho ions provide additional connectivity in the magnetic sublattice. In fact, this sort of transformation of the connectivity of a frustrated system has not, to our knowledge, been previously examined in any geometrically frustrated magnetic material and would be structurally impossible in water ice. Given the significant change in topology, one might expect the low temperature spin ice behavior to be entirely disrupted. We measured the specific heat as well as the ac and dc magnetic susceptibilities of this series, with the intention of investigating how the ice-like properties (e.g., the dynamic freezing and the entropy deficit) are altered by enhancing the connectivity.

The initial characterization of the samples came from the dc magnetization, which was measured both as a function of temperature and magnetic field. Field dependent data for all of the samples at $T = 2$ K demonstrated that the moment saturated at 5 $\mu_B$/Ho, which is about half of the theoretically expected value. This is consistent with an Ising-like character for all of the moments, and is somewhat surprising, given that the oxygen environments of the R and M sublattices should be different. Constant field $M(T)$ data taken for $T = 10 - 20$ K at $H = 0.1$ T were fit to the Curie-Weiss formula in order to determine the average mean field spin-spin interaction through the Weiss temperature, $\theta_W$. We find that $\theta_W$ is positive (ferromagnetic) for $x = 0$, as demonstrated previously,[3] but decreases with increasing $x$ as shown in Fig. 2. In fact, $\theta_W$ changes sign between $x = 0.3$ and $x = 0.4$, indicating that the predominant spin-spin interactions become antiferromagnetic for larger values of $x$.    This could imply either that the alteration of the lattice is changing the exchange and dipole interactions between the Ho spins on the R sites, or that the additional spins on the M sites have antiferromagnetic interactions with each other or with the spins on the R sites, but the spin-spin interactions are clearly changed by the stuffing of additional Ho into the lattice.

We plot the measured specific heat data, $C(T)$, in Fig. 3a, and show the fits used to subtract out the lattice contribution.  These fits were obtained from scaling polynomial fits to heat capacity measurements on the structurally similar non-magnetic materials $Lu_2(Ti_{2-x}Lu_x)O_{7-x/2}$, with $x = 0$, 0.3, and 0.67.  The magnetic specific heat was then obtained by subtracting out the lattice contribution and the low temperature Schottky peak associated with hyperfine contributions, using the fit obtained by previous workers.[16]  The magnetic entropy, $S(T)$, at $H = 0$ and $H = 1$ T was then determined by integrating $C_{magnetic}(T)/T$, as shown in Fig. 3c.  This plot assumes that $S = 0$ at $T = 0$, and would extrapolate to $S(T) = R\ln2$ at high temperatures for a system of Ising spins which explored all possible states.  For $x = 0$ and $H = 0$, we obtained an integrated entropy at high temperatures close to $S = R[\ln2 - \frac{1}{2}\ln(^3/_2)]$, in excellent agreement with previous work.[16,25]  The application of a magnetic field restored some of the "missing" entropy as first reported by Ramirez et al.,[4] an effect that is attributed to the decrease in the ground state degeneracy when the system's symmetry is broken by the applied field. Remarkably, as $x$ is increased, the integrated magnetic entropy at $H = 0$ is essentially constant, and the effect of applied field (e.g., the difference between the integrated entropy at $H = 0$ and $H = 1$ T) is noticeably diminished.  This implies that the more highly connected spins on the stuffed lattices show the same total magnetic entropy as the spin-ice state, despite the increased connectivity and average antiferromagnetic interactions.  The presence of a zero point entropy in other Ising systems is theoretically expected,[6] but it is rather astonishing that the same amount of zero point entropy is retained with the increased connectivity associated with stuffing. The limited effect of an applied field on the stuffed material suggests that stuffing the lattice actually makes the zero point entropy more robust.

The temperature dependence of ac susceptibility ($\chi_{ac}$) is shown in Fig. 4a, which clearly shows that frequency dependence develops below $T = 3$ K in the $x = 0$ material. This is consistent with the susceptibility peak seen in previous reports on spin ice materials and associated with dynamic freezing to the low temperature state. [17,26] The magnitude of the susceptibility at the lowest temperatures decreases with increasing $x$, which can be attributed to the change in $\theta_W$, but there is also a decrease in the frequency dependence of $\chi_{ac}$ as is illustrated in Fig. 4b. The $\chi_{ac}(f)$ data show a monotonic decrease with increasing frequency for all of the samples, as is expected for dynamic spin freezing, but the decrease in $\chi_{ac}(f)$ becomes smaller and occurs at higher frequency for larger values of $x$. This trend with increasing Ho content indicates that the spin relaxation time is shorter for larger $x$, i.e., that the spins relax more quickly and thus may freeze only at lower temperatures. This is again quite surprising, given that stuffing increases connectivity and might be expected to have a higher propensity toward freezing. Indeed, spin freezing in other geometrically frustrated materials can be suppressed by diluting the frustrated lattice through non-magnetic ion substitution on the magnetic sites (effectively reducing the connectivity).[27]

The above results are striking, since stuffing transforms the magnetic lattice from a well ordered system of corner-sharing tetrahedra, to a disordered system of side-sharing tetrahedra and also causes a switch from predominantly ferromagnetic to predominantly antiferromagnetic interactions. Furthermore, dynamic spin freezing appears suppressed with stuffing of the lattice, suggesting that the low lying spin states are even closer in energy than in ordinary spin ice. These data provide the first experimental evidence that the zero point entropy is a feature of a broader range of frustrated Ising spin systems. That the zero point entropy retains the same value and becomes more robust against applied field in the stuffed systems may be evidence that the "ice rules" have relevance beyond the pure spin ice system. Perhaps more importantly, the successful synthesis of this system paves the way for new studies of all of the geometrically frustrated pyrochlore compounds as a function of increased connectivity, and thus opens a new window on arguably the most important model system of geometrical magnetic frustration.

## METHODS

Polycrystalline samples of $Ho_2(Ti_{2-x}Ho_x)O_{7-x/2}$ were synthesized by heating the stoichiometric amounts of $Ho_2O_3$ (Cerac, 99.9%) and $TiO_2$ (Cerac, 99.9%) at 1700 $^o$C in a static argon atmosphere for 12 hours. The powders were intimately mixed, pressed into pellets and wrapped in molybdenum foil before heating. The argon atmosphere was achieved in a vacuum furnace first evacuated to about $10^{-8}$ torr and then back-filled with argon (Airgas, 99.9%) to room pressure.

The crystal structures were characterized through x-ray powder diffraction data using CuK$\alpha$ radiation and a diffracted beam monochromator. Structural refinements were made using the Bruker AXS software package TOPAS 2.1$^{\copyright}$ operated with a Pseudo-Voight TCHZ fitting profile. Ho and Ti occupancies were allowed to refine freely on both the R and M sites of the pyrochlore, with the constraint that their total occupancies from both sites maintained the correct stoichiometric ratios. The small non-zero Ti occupancy on the R site for stoichiometric $Ho_2Ti_2O_7$ and neighboring compositions is taken as zero within experimental uncertainty, with the Ho/Ti mixing on the R site seen to start at x = 0.4.

Magnetic and specific heat measurements were performed on pressed pellets in Quantum Design PPMS and MPMS cryostats. Heat capacity measurements used a standard semi-adiabatic heat pulse technique, and the addendum heat capacity was measured separately and subtracted. The ac susceptibility was measured using an ac excitation field of 0.5 Oe. All the samples for susceptibility were cut to needle-like shapes and were oriented along the direction of the applied field(s) in order to minimize demagnetization effects.

**Figure Captions**

**Figure 1.** Structural details of the materials studied: $Ho_2(Ti_{2-x}Ho_x)O_{7-x/2}$. a) The cubic pyrochlore ($R_2M_2O_7$) lattice. The corner-sharing tetrahedral sublattices of Ho and Ti are shown in the $Ho_2Ti_2O_7$ structure (left), with the outline in green depicting edge-sharing tetrahedral interactions that develop as extra Ho replaces Ti atoms. At x = 0.67, Ho and Ti reside randomly on the same sites, forming the fluorite phase (right). b) Main panel: lattice parameter and fraction of R-site disorder (occupancy of Ti on the R-site of the pyrochlore structure) in the stuffed material are plotted as a function of $x$. Lines drawn are guides to the eye. Inset: metal-metal coordination around Ho (red) and Ti (blue) in $Ho_2Ti_2O_7$ with only R and M sites shown; oxygen atoms are omitted. Lines in bold are drawn to highlight segments of the original corner-sharing tetrahedral motif. Purple atom represents a possible Ti site that can be "stuffed" with extra Ho.

**Figure 2.** Temperature dependence of the magnetization. (a) The Weiss temperature ($\theta_W$) plotted versus $x$ for the various $Ho_2(Ti_{2-x}Ho_x)O_{7-x/2}$ samples, determined from fits to the Curie-Weiss Law. Note the crossover from positive to negative $\theta_W$ with increasing $x$. (b) The inverse dc magnetic susceptibility versus temperature for $x = 0$ and 0.67 illustrating the crossover in $\theta_W$ which is taken from the intercept with the horizontal axis (lines are fits to the data used to determine $\theta_W$, as described in the text; the effective moment obtained from these fits is within 2% of the accepted value for Ho.).

**Figure 3.** Thermal characterization of stuffed spin ice. (a) The total specific heat of $Ho_2(Ti_{2-x}Ho_x)O_{7-x/2}$ for $x = 0$, 0.3 and 0.67 at $H = 0$ and 1 T. The dashed and solid lines

represent the lattice and nuclear contributions, as described in the text.  (b) The total magnetic entropy (integrated from below $T = 1$ K to $T = 22$ K) as a function of $x$ at $H = 0$ and 1 T. The reproducibility of the integrated entropy was within ± 5% from run to run, and the total uncertainty somewhat larger (±10%) due to uncertainty in the subtraction of the lattice contribution to the specific heat.  The blue and black dashed lines represent the predicted ice entropy and total spin entropy values, respectively.  (c) The temperature dependence of the integrated entropy for our limiting compositions, at $H = 0$ and 1 T.

**Figure 4.**    The low temperature ac magnetic susceptibility.    (a) The $H = 0$ ac susceptibility of $Ho_2(Ti_{2-x}Ho_x)O_{7-x/2}$, for three values of $x$.  The frequency dependent peak around $T = 2$ K, usually associated with dynamic freezing into a disordered state governed by the ice rules, was seen to diminish with increasing $x$.  (b) The frequency dependence of the ac susceptibility at $T = 2$ K and $H = 0$ spanning the range of $x$ studied (data are normalized to the $f = 10$ kHz values).

ACKNOWLEDGEMENTS

The authors gratefully acknowledge financial support from the National Science Foundation and helpful discussions with R. Moessner and A. P. Ramirez.  R.S.F. thanks the CNPq-Brazil for sponsorship.

Figure 1.

a)

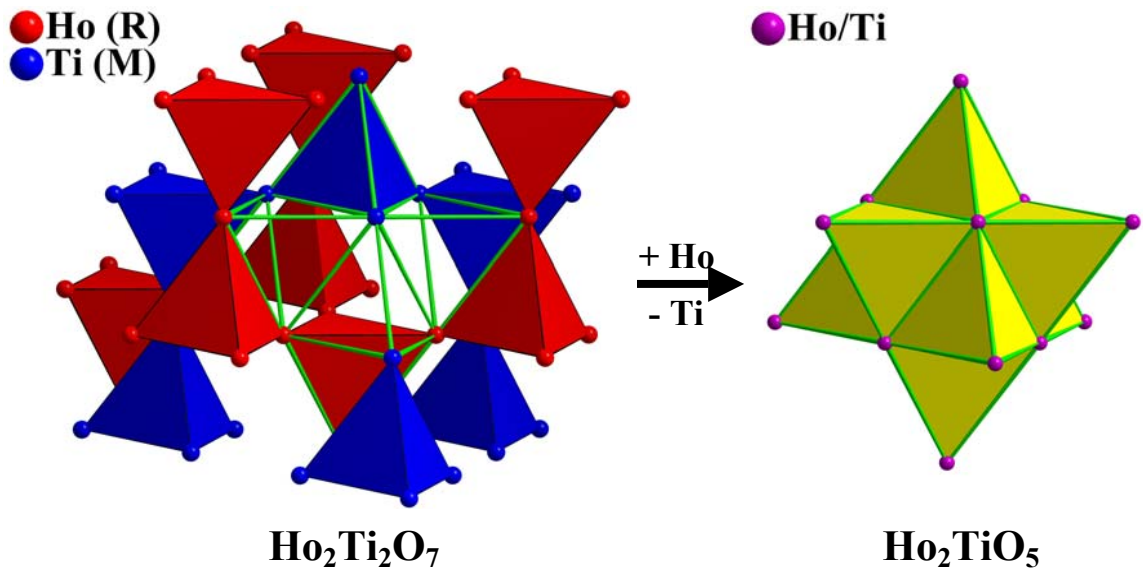

b)

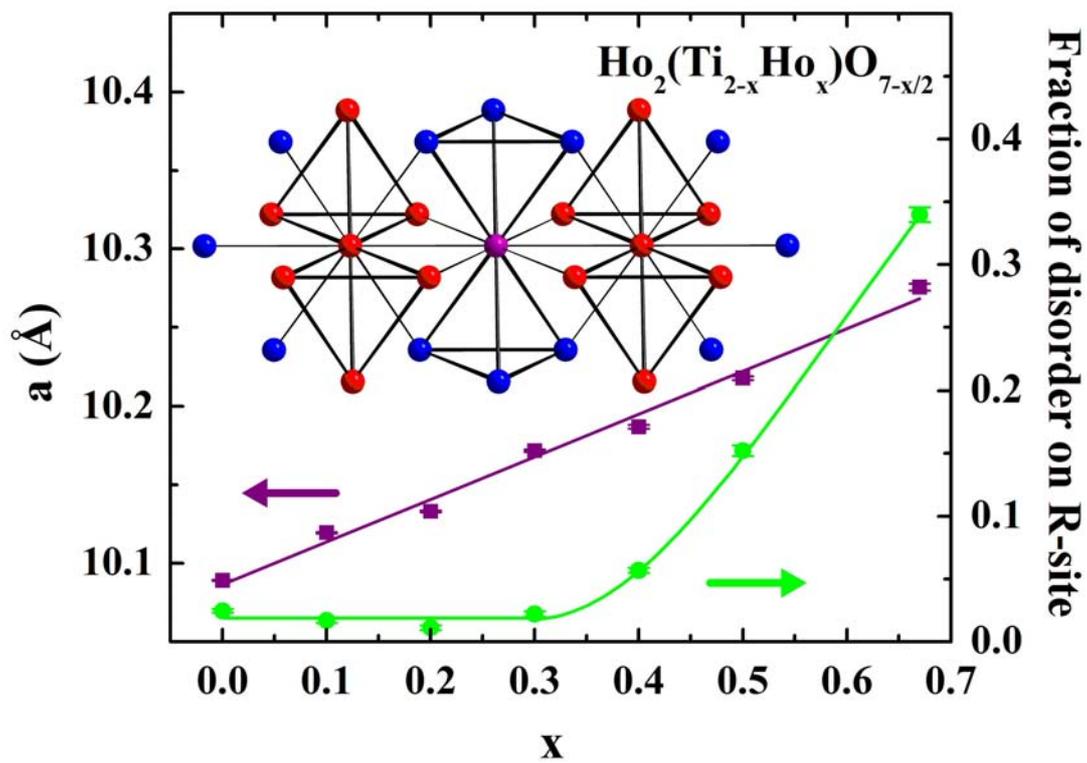

Figure 2.

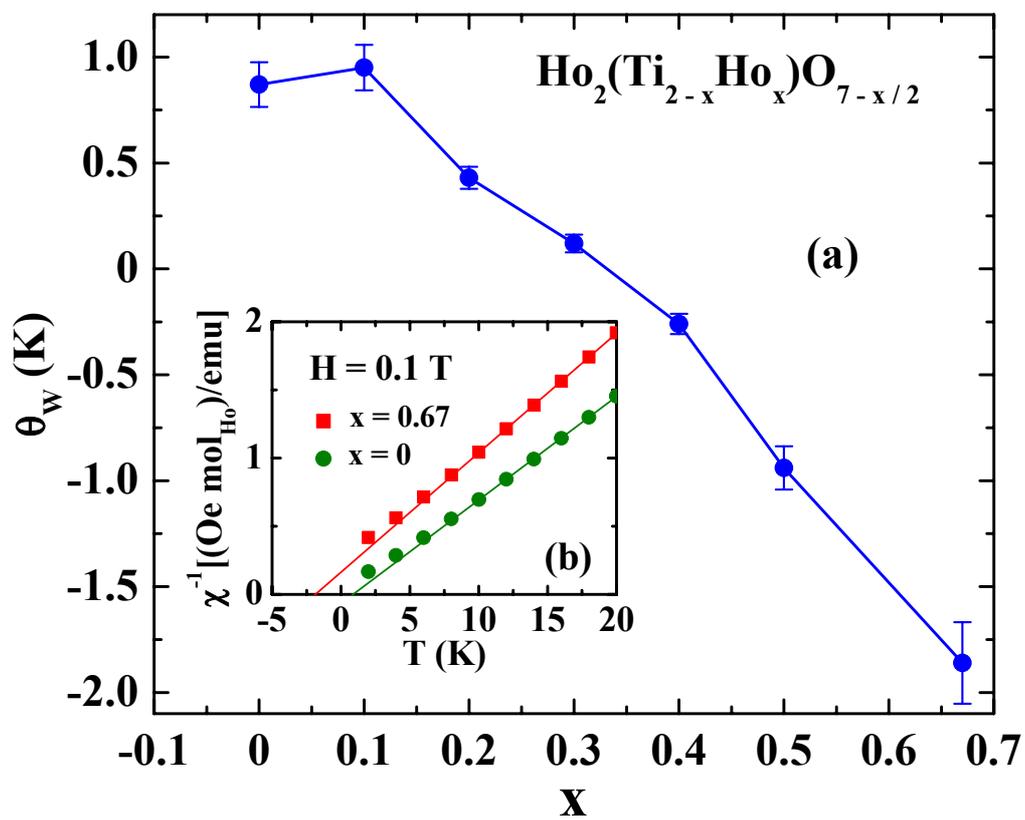

Figure 3.

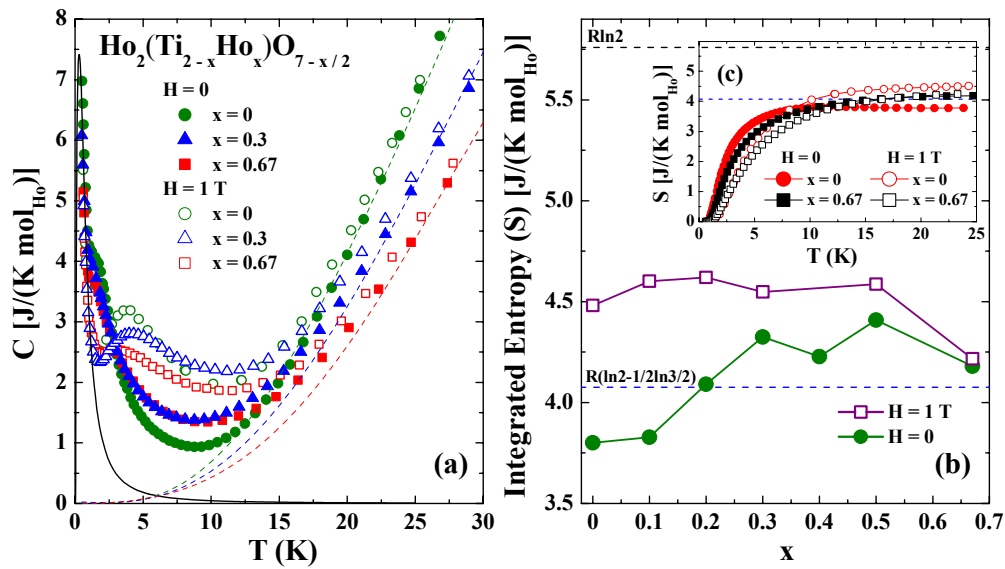

**(a)** Ho$_2$(Ti$_{2-x}$Ho$_x$)O$_{7-x/2}$

H = 0
- x = 0
- x = 0.3
- x = 0.67

H = 1 T
- x = 0
- x = 0.3
- x = 0.67

**(b)**

- H = 1 T
- H = 0

Rln2

R(ln2-1/2ln3/2)

**(c)**

H = 0
- x = 0
- x = 0.67

H = 1 T
- x = 0
- x = 0.67

Figure 4.

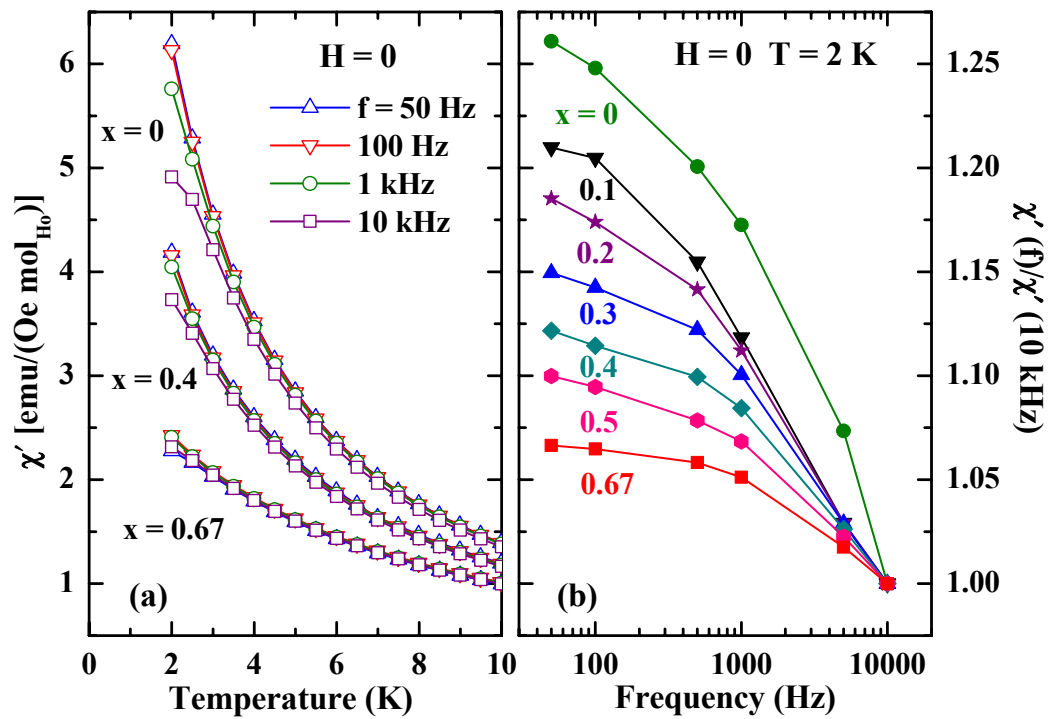

**References**

————————————————

1. Bernal, J. D. and Fowler, R. H. A theory of water and ionic solution, with particular reference to hydrogen and hydroxyl ions. *J. Chem. Phys.* **1**, 515–548 (1933).

2. Pauling, L. The structure and entropy of ice and of other crystals with some randomness of atomic arrangement. *J. Am. Chem. Soc.* **57,** 2680-2684 (1935).

3. Harris, M. J., Bramwell, S. T., McMorrow, D. F., Zeiske, T., and Godfrey, K. W. Geometrical frustration in the ferromagnetic pyrochlore $Ho_2Ti_2O_7$. *Phys. Rev. Lett.* **79** 2554-2557 (1997).

4. Ramirez, A. P., Hayashi, A., Cava, R. J., and Siddharthan, R. Zero-point entropy in 'spin ice.' *Nature* **399,** 333-335 (1999).

5. Bramwell, S. T. and Gingras, M. J. P. Spin ice state in frustrated magnetic pyrochlore materials. *Science* **294**, 1495-1501 (2001).

6. Bramwell, S. T., Gingras M. J. P., and Holdsworth, P. C. W. in *Frustrated Spin Systems* (ed. Diep, H. T.) 367-456 (World Scientific, Singapore, 2004).

7. Higashinaka, R., Fukazawa, H., and Maeno, Y. Anisotropic release of the residual zero-point entropy in the spin ice compound $Dy_2Ti_2O_7$: Kagome ice behavior. *Phys. Rev. B* **68**, 014415 (2003).

8. Ramirez, A. P. in *Handbook of Magnetic Materials* Vol. 13 (ed. Buschow, K. J. H.) 423-520 (Elsevier Science, Amsterdam, 2001).

9. Moessner, R. Magnets with strong geometric frustration. *Can. J. Phys.* **79**, 1283-1294 (2001).

10. Gaulin, B. D., and Gardner, J. S., in *Frustrated Spin Systems* (ed. Diep, H. T.) 457-490 (World Scientific, Singapore, 2004).

11. Raju, N. P., Dion, M., Gringras, M. J. P., Mason, T. E., and Greedan, J. E. Transition to long-range magnetic order in the highly frustrated insulating pyrochlore antiferromagnet $Gd_2Ti_2O_7$. *Phys. Rev. B* **59**, 14489-14498 (1999).

12. Ramirez, A. P. *et al*. Multiple field-induced phase transitions in the geometrically frustrated dipolar magnet: $Gd_2Ti_2O_7$. *Phys. Rev. Lett.* **89**, 067202 (2002).

13. Champion J. D. M. *et al*. $Er_2Ti_2O_7$: evidence of quantum order by disorder in a frustrated antiferromagnet. *Phys. Rev. B* **68**, 020401 (2003).

14. Hodges, J. A. *et al*. First-order transition in the spin dynamics of geometrically frustrated $Yb_2Ti_2O_7$. *Phys. Rev. Lett.* **88**, 077204 (2002).

15. Gardner, J. S. *et al*. Cooperative paramagnetism in the geometrically frustrated pyrochlore antiferromagnet $Tb_2Ti_2O_7$. *Phys. Rev. Lett.* **82**, 1012-1015 (1999).

16. Bramwell, S. T. *et al*. Spin correlations in $Ho_2Ti_2O_7$: a dipolar spin ice system. *Phys. Rev. Lett.* **87**, 047205 (2001).

17. Matsuhira, K., Hinatsu, Y., Tenya, K., and Sakakibara, T. Low temperature magnetic properties of frustrated pyrochlore ferromagnets $Ho_2Sn_2O_7$ and $Ho_2Ti_2O_7$. *J. Phys. Condens. Matter* **12**, L649 – L656 (2000).

18. Fukazawa, H., Melko, R. G., Higashinaka, R., Maeno, Y., and Gingras, M. J. P. Magnetic anisotropy of the spin-ice compound $Dy_2Ti_2O_7$. *Phys. Rev. B* **65**, 054410 (2002).

19. Snyder, J. *et al*. Low-temperature spin freezing in the $Dy_2Ti_2O_7$ spin ice. *Phys. Rev. B* **69**, 064414 (2004).

20. Wiebe, C. R. *et al*. Frustration-driven spin freezing in the S=1/2 fcc perovskite $Sr_2MgReO_6$. *Phys. Rev. B* **68,** 134410 (2003).

21. Karunadasa, H., Huang, Q., Ueland, B. G., Schiffer, P., and Cava, R. J. $Ba_2LnSbO_6$ and $Sr_2LnSbO_6$ (Ln = Dy, Ho, Gd) double perovskites: lanthanides in the geometrically frustrating fcc lattice. *Proc. Natl. Acad. Sci. U.S.A.* **100**, 8097-8102 (2003).

22. Subramanian M. A., Aravamudan G., and Rao G. V. S. Oxide pyrochlores – a review. *Prog. Sol. State Chem.* **15**, 55-143 (1983).

23. Sukhanova, G. E., Guseva, K. N., Kolesnikov, A. V., and Shcherbakova, L. G. Phase-equilibria in the $TiO_2$-$Ho_2O_3$ system. *Inorg. Mater.* **18,** 1742-1745 (1982).

24. Shamrai, G. V., Magunov, R. L., Sadkovskaya, L. V., Stasenko, I. V., and Kovalevskaya, I.P. The system $Ho_2O_3$-$TiO_2$. *Inorg. Mater.* **27,** 140-141 (1991).

25. Cornelius, A. L. and Gardner, J. S. Short-range magnetic interactions in the spin-ice compound $Ho_2Ti_2O_7$. *Phys. Rev. B* **64**, 060406 (2001).

26. Matsuhira, K., Hinatsu, Y., and Sakakibara, T. Novel dynamical magnetic properties in the spin ice compound $Dy_2Ti_2O_7$. *J. Phys. Condens. Matter* **13**, L737-L746 (2001).

27. Ramirez, A. P., Espinosa, G. P., and Cooper, A. S. Strong frustration and dilution-enhanced order in a quasi-2D spin glass. *Phys. Rev. Lett.* **64**, 2070-2073 (1990).